\documentclass[11pt,a4paper]{llncs}
\usepackage{amsmath}
\usepackage{amssymb}
\usepackage{graphicx}
\usepackage{marvosym}
\usepackage{url}
\usepackage{fancyhdr}
\usepackage{enumitem}

\usepackage{geometry}
\newcommand{\keywords}[1]{\par\addvspace\baselineskip
\noindent\keywordname\enspace\ignorespaces#1}

\fancypagestyle{firstpage}{\fancyhf{}
}

\begin{document}

%\mainmatter  % start of an individual contribution

\title{\LARGE{Programmable Data Planes for Network Security}}

%\titlerunning{Lecture Notes in Computer Science: Authors' Instructions}


\author{Gursimran Singh\textsuperscript{1}, H.B. Acharya\textsuperscript{2}, Minseok Kwon\textsuperscript{1}}
%{\large{Afg Cer  \and A. Bcd \and P. Qwe \and R. Bcgfee \and R.Y. Tedf}}
\institute{\textsuperscript{1}Rochester Institute of Technology, \textsuperscript{2}Oklahoma State University}
%{\large{School of XYZ and RST, University of ABCDEF,\\ Qwergh-12345, XYZ}}

%\author{Alfred Hofmann%
%\thanks{Please note that the LNCS Editorial assumes that all authors have used
%the western naming convention, with given names preceding surnames. This determines
%the structure of the names in the running heads and the author index.}%
%\and Ursula Barth\and Ingrid Haas\and Frank Holzwarth\and\\
%Anna Kramer\and Leonie Kunz\and Christine Rei\ss\and\\
%Nicole Sator\and Erika Siebert-Cole\and Peter Stra\ss er}

%\institute{Springer-Verlag, Computer Science Editorial,\\
%Tiergartenstr. 17, 69121 Heidelberg, Germany\\
%\url{http://www.springer.com/lncs}}


%\toctitle{Lecture Notes in Computer Science}
%\tocauthor{Authors' Instructions}

\maketitle

\thispagestyle{firstpage}

\begin{abstract}
The emergence of programmable data planes, and particularly switches supporting the P4 language, has transformed network security by enabling customized, line-rate packet processing. These switches, originally intended for flexible forwarding, now play a broader role: detecting and mitigating attacks such as DDoS and spoofing, enforcing next-generation firewall policies, and even supporting in-network cryptography and machine learning. These capabilities are made possible by techniques such as recirculate-and-truncate and lookup-table precomputation, which work around architectural constraints like limited memory and restricted instruction sets.

In this paper, we systematize recent advances in security applications built on programmable switches, with an emphasis on the capabilities, challenges, and architectural workarounds. We highlight the non-obvious design techniques that make complex in-network security functions feasible despite the constraints of the hardware platform, and also comment on remaining issues and emerging research directions.

\end{abstract}

%\begin{abstract}
%The abstract should summarize the contents of the paper and should
%contain at least 70 and at most 150 words. It should be written using the
%\emph{abstract} environment.
%\keywords{We would like to encourage you to list your keywords within
%the abstract section}
%\end{abstract}

\section{Introduction}

Computer networks face increasingly sophisticated attacks, which require rapid and adaptive response. Conventional networking equipment, developed as a closed design, limits network operators to use facilities hard-coded by  manufacturers. The first step in removing such limitations was the development of Software-Defined Networking (SDN), which centralizes control plane functions and allows for flexibility in controller-level applications (with a standard south-bound interface between controller and switch) \cite{mckeown2008openflow}. The second step has been the rise of programmable network switches, which allow the operator to carry out more and more functions in the data plane, i.e. on the switch itself.

Programmable switches empower network administrators to implement customized packet processing functions, in-line in the switch. This has strong implications for speed and scale (as sending things to the controller almost inevitably leads to a bottleneck and prevents line-rate packet processing). Programming Protocol-independent Packet Processors (P4) has emerged as the \emph{de facto} data plane programming language, enabling the deployment of sophisticated packet processing algorithms in network switches. This includes not only network management and telemetry, but also network security, making programmable switches a compelling platform for deploying in-fabric defenses. This paper surveys these emerging security applications. 

\subsection{P4 and Programmable Switches}

P4 \cite{bosshart2014p4} is a domain-specific language that allows the network administrator (not the router manufacturer) to specify the structure of packets and control how these packets are processed by the data plane of network devices. Its aim was to allow the network admin to customize the processing pipeline for packets belonging to any free-form protocol (hence the name, programming protocol-independent packet processors). The Portable Switch Architecture (PSA) and Protocol Independent Switch Architecture (PISA) describe the capabilities of switches that can be programmed using P4, including a programmable parser, match-action pipeline, and deparser \cite{p4_arch_spec}.

Programmable switches, like all switches, operate on the match-action paradigm. A switch matches a field from a packet against values in a table, and the matched entry specifies corresponding actions to apply to the packet. These switches can operate at line rates (100 Gbps or higher), so their logic units execute simple operations only, primarily bit manipulation and basic arithmetic. The pipeline architecture consists of multiple stages, containing memory elements (SRAM, TCAM) and ALUs, and also a parser and deparser.

The main difference from other switches is that the parser, which extracts fields from the packet, is \emph{programmable}. In other words, the end user can specify a schema for a new protocol, and the switch will slice up the packet according to this schema. This allows the flexibility to add support for new protocols at any time, and have it be supported by any switch that adheres to the P4 standard. Finally, the deparser stitches the packet back together after processing, so it can be forwarded to its destination, looped back, mirrored, etc.

A couple of intentional limitations in the architecture, for the sake of fast packet processing, cause the programmable switch to be limited as an  "active network" node.
\begin{enumerate}[leftmargin=*]
    \item The parser is not truly a top-down parser, able to parse the packet as a general-purpose compiler parses code. It is much more limited and essentially fetches bit-slices of a given length starting from a given offset. So complex parsing (as required by say HTTP, with its optional, re-ordered, and variable-length fields) is not supported. In other words, the language offers sufficient flexibility for defining novel protocol headers and tags, but not enough for full in-network processing (for example, for query processing in the network fabric).

    \item Complex operations, like floating-point operations or encryption, are not natively supported by standard P4-compliant architectures. This, again, limits the intelligence of the switch when the majority of traffic is encrypted, and the power of in-network computation.
\end{enumerate}

We now come to the insight that motivated this paper. Programmable switches are quite limited as a platform for general computation, but are still a very attractive platform for network security applications. Most importantly, they can process packets extremely fast, on the order of 100x faster than a typical CPU~\cite{jepsen2019fast}. They do have some instructions for programming, using state associated with packets (registers and monitors), though limited to operations that were thought to be necessary for protocol- and target-independent network operations. But in addition they have significant chip area for memory to support tables (early switches, like Reconfigurable Match Tables (RMT), allocated over 50\% of its chip area for memory \cite{bosshart2013forwarding}) and such tables can be creatively used as look-up tables for computation, not just to match and forward packets.  

This has led to two important research directions -- the first, directly using such switches to provide security applications such as DDoS mitigation and firewalls; the other, ``hacking" the architecture by using lookup tables for encryption~\cite{chen2020implementing} or the recirculate-and-truncate technique packets to achieve true parsing and pattern-matching~\cite{jepsen2019fast}, i.e. side-stepping the limitations mentioned above. Accordingly, we present a comprehensive study that covers both straightforward (``kosher'') and workaround-based (``hacky'') approaches for implementing network security applications in the data plane, as well as a discussion of the open challenges and research directions.

This paper has the following main contributions:
\begin{itemize}[leftmargin=*]
    \item We systematize programmable data plane security research across three major categories (\textbf{For in-network attack mitigation}, \textbf{As middleboxes and firewalls}, and \textbf{For in-network computations and ML-based security}), and analyzing fundamental trade-offs between programmability, performance, and security.

    \item We highlight the non-obvious design techniques (such as \textbf{recirculate-and-truncate} and \textbf{lookup-table pre-computation}) that make complex security functions feasible despite hardware constraints, and identify current challenges and future research directions in the field.
\end{itemize}

The rest of this paper is organized as follows. Section II explores programmable switches as ``better switches'', i.e. enhanced packet-forwarding devices and how they help with threat mitigation in networks. Section III expands on their role as security middleboxes, notably  application-layer firewalls and cryptography. Section IV then discusses their emerging use as platforms for machine learning-based security. We then present some discussion, focusing on broad challenges and future directions, and conclude the paper.

\section{Smart Switches: In-Network Attack Mitigation}

In this section, we focus on the network security role of programmable switches that act simply as an updated, more powerful version of traditional switches. Even in this limited role, such switches can detect and mitigate many attacks, some of them at line rate (providing superior performance compared to traditional software-based solutions, and much lower costs than dedicated hardware). We first mention P4-based solutions to address two traditional network threats, spoofing and DDoS attacks, then continue on to discuss how these switches also help with hardening the network and making it  quickly adaptable to threats.

\subsection{Spoofing Protection}

Spoofing attacks, where attackers impersonate legitimate devices, have always been a persistent threat in network environments, and traditional switches have defended against such attacks for decades. CloudFlare reports that even now, almost all large-scale layer-3 DDoS attacks rely on IP spoofing \cite{majkowski2016realcause}. Several anti-spoofing mechanisms have been developed with programmable switches, essentially updating the tools found in traditional L2 switches (Port Security, DHCP snooping, Dynamic ARP Inspection, and IP Source Guard).

\begin{itemize}[leftmargin=*]
    \item \textbf{NetHCF}: Li et al. \cite{li2019nethcf} propose NetHCF, which maintains a mapping table between IP addresses and hop counts (IP-to-Hop-Count or IP2HC), allowing it to infer spoofed traffic by comparing hop counts in arriving packets with expected values from legitimate sources. The IP2HC table is stored in-controller with the most active IP entries cached in the data plane using registers. aggregating nodes with common IP prefixes and hop-count values.

    \item \textbf{DroPPPP}: Simsek et al. \cite{simsek2019dropppp} develop DroPPPP, which detects spoofed addresses by calculating a hash of the source IP and MAC addresses, then comparing it against previously stored hash values. DroPPPP uses three registers (hash values, timestamps, attack flag) and identifies and blocks spoofed packets at line rate. 

    \item \textbf{P4DAD}: Kuang et al. \cite{kuang2020p4dad} proposed P4DAD to secure Duplicate Address Detection. P4DAD stores bindings between IPv6 addresses and ports in the switch (much like a DHCP binding table), and verifies the authenticity of Neighbor Advertisement messages, preventing attackers from disrupting the address configuration process.

    \item Other implementations, such as those by Narayanan et al. \cite{narayanan2019mitigation} and Gondaliya et al. \cite{gondaliya2020comparative}, have explored additional anti-spoofing techniques including Network Ingress Filtering (NIF), Reverse Path Forwarding (RPF), Spoofing Prevention Method (SPM), and Source Address Validation Improvement (SAVI) solution for DHCP.
\end{itemize}

P4-based anti-spoofing mechanisms offer much lower latency and full network-wide visibility of all traffic, compared to host-based approaches. They also do not incur the costs and delays of additional hardware or modifications to routers, as the required modifications to the switch behavior are made through dataplane programming (not hardware-level changes).

\subsection{DDoS Attack Mitigation}

A second traditional area of network security involves Distributed Denial-of-Service attacks. Programmable switches have been used to develop robust defenses that balance two goals: high throughput and accuracy.

P4-based algorithms were quickly developed to detect heavy hitters (traffic flows of large volume), which often constitute DDoS attacks. For example, \textbf{HashPipe} \cite{sivaraman2017heavy} identifies the k-heaviest flows with high accuracy entirely in the data plane using a pipeline of hash tables.
 \textbf{PRECISION} \cite{benbasat2018efficient} uses probabilistic recirculation to monitor a small fraction of packets for a second pipeline traversal, balancing memory constraints with accuracy. \textbf{MV-Sketch} \cite{tang2020fast} implements an invertible sketch that applies majority vote algorithms to find candidate heavy flows. And \textbf{Elastic Trie} \cite{kucera2018seek} targets hierarchical heavy hitters, changes in traffic patterns, and superspreaders, which can be used for DDoS, anomaly, worm, and spam detection.

Subsequently, P4-based systems have been built to detect and respond to such attacks.
\begin{itemize}[leftmargin=*]
\item \textbf{POSEIDON} \cite{zhang2020poseidon}: Provides network operators with an expressive policy language to specify DDoS attack mitigation strategies. POSEIDON partitions functions between P4 switches and general-purpose servers, adapting to dynamic attacks by generating new defense policies and reconfiguring switches.

\item \textbf{Jaqen} \cite{liu2021jaqen}: Implements switch-native DDoS detection and mitigation in the data plane, covering a wide range of volumetric attacks within ISPs, and provides APIs for querying metrics and flexible mitigation.

\item \textbf{DDoSD-P4} \cite{lapolli2019offloading}: Implements a fully in-network inspection mechanism with three stages: entropy estimation, traffic characterization, and anomaly detection. \textbf{EUCLID} \cite{silveira2020euclid} extends DDoSD-P4 to enable scalable detection and mitigation of volumetric DDoS attacks in the data plane.

\item \textbf{INDDoS} \cite{ding2021innetwork}: Employs a novel data structure called BACON (combining Bitmaps and Count-Min Sketch) to estimate per-destination flow cardinality, i.e. they identify hosts to which a large number of flows are targeted as likely DDoS victims.
\end{itemize}

There also exist implementations that focus on specific protocols and attack vectors commonly used in DDoS.
\begin{itemize}[leftmargin=*]
\item \textbf{DNS Amplification Defense}: Khooi et al. \cite{khooi2020dida} proposed DIDA, a distributed in-network defense architecture against DNS amplification attacks, monitoring both requests and responses across the network.

\item \textbf{SIP DDoS Defense}: Febro et al. \cite{febro2019distributed}  detect DDoS attacks using the Session Initiation Protocol by monitoring INVITE and REGISTER packets at the first-hop switch.
\end{itemize}

These DDoS mitigation techniques are not only much faster than software-based solutions (or even other SDN/NFV defenses that rely on the control plane), but unlike traditional traffic scrubbing centers that use expensive proprietary hardware (say \$ 1 M -- \$10 M), they combine high performance and simple reconfigurability at a very affordable price (\$5 k -- \$10 k for an industrial-scale switch).

\section{Smart Switches as Middleboxes and Firewalls}

Programmable switches not only act as enhanced switches for in-line threat mitigation, they also increasingly serve the functions previously performed by full-featured middleboxes. In this section we examine systems that build such switches into firewalls (including application-layer firewalls) or cryptographic endpoints (say, for a VPN) directly in the network fabric. 

\subsection{Firewalls and Access Control}

Firewalls control traffic flow between networks based on predetermined security rules, either filtering in the network layer (typically using the header fields: source IP, source port, destination IP, destination port, and protocol) or in the application layer (Deep Packet Inspection -- for example, URL filtering). Programmable data planes offer new opportunities for implementing high-performance, flexible firewall functionality directly in network hardware, as seen in the following systems:
\begin{itemize}[leftmargin=*]
    \item \textbf{P4Guard}: Datta et al. \cite{datta2018p4guard} developed P4Guard, a configurable firewall using P4. The authors define a high-level, hardware-independent security policy language; the controller translates these policies written in the language, into match-action table rules that are deployed to the switch. P4Guard implements tables for firewall rules, based on IP, TCP, UDP, and ARP header fields, and maintains a counter table to record statistical flows, periodically sending this information to the controller for analysis.

    \item \textbf{CoFilter}: Cao et al. \cite{cao2018cofilter} proposed a stateful packet filter implemented on a switch. CoFilter maintains flow state by computing a hash of the 5-tuple and storing it as a flow ID (fid), enabling stateful filtering with limited memory. CoFilter uses a management server to update flow tables, to avoid hash collisions over time.

    \item \textbf{Stateful Firewalls}: Li et al. \cite{li2020sdn} implemented a stateful firewall for cloud environments using P4. Their approach leverages the expressiveness of P4 to parse packets and filter them based on firewall policies while tracking TCP connections to allow only packets belonging to established sessions.

    \item \textbf{Authentication and Port Knocking}: ``port knocking'' is a covert channel where a signal is sent by sending requests to a specific sequence of ports. Almaini et al. \cite{almaini2019delegation} implemented port knocking on programmable switches as an authentication mechanism. The system stores authenticated and unauthenticated connections using match-action table rules inserted by the controller. In a follow-up work \cite{almaini2021lightweight}, they add One-time Password (OTP) authentication to protect against replay attacks.

    Zaballa et al. \cite{zaballa2020p4knocking} presented multiple implementations of port knocking, ranging from pure data-plane implementations (using registers to store knocking states), to hybrid approaches that delegate some functions to the controller.

    \item \textbf{Mobile Network Firewalls}: Ricart-Sanchez et al. \cite{ricart2018hardware} developed a firewall for 5G mobile network infrastructure, positioned between edge and core networks. Their design was later extended \cite{ricart2019netfpga} to support multi-tenant 5G infrastructures.
\end{itemize}

P4-based firewalls offer several advantages over traditional header-based (network-layer) firewalls. They are just as fast as a hardware firewall, and rapidly adapt to new threats without hardware upgrades, unlike traditional fixed-function firewalls. More importantly, however, the flexibility of a programmable dataplane allows for the development of more general security systems, combining network security and management. We mention a few important systems below.
\begin{itemize}[leftmargin=*]
    \item \textbf{NetWarden}: Xing et al. \cite{xing2020netwarden} created a defense mechanism that preserves TCP performance while mitigating covert storage and timing channels. NetWarden performs header inspection and modification on fields commonly used in covert channels (e.g., TTL, TCP reserved fields). For advanced storage channel protection, it replaces headers with newly generated ones and stores mappings between original and new headers in stateful memory.

    \item \textbf{Poise}: Kang et al. \cite{kang2020programmable} developed a system for context-aware security policies, particularly for Bring Your Own Device (BYOD) environments. Poise translates high-level policies into P4 programs that can enforce dynamic access control based on device runtime context (e.g., operating system version).

    \item \textbf{FastFlex}: Xing et al. \cite{xing2019architecting} created an architectural framework supporting various defense mechanisms. Their system transforms defense applications (``boosters") into Packet Processing Modules that can be optimally mapped to network resources, enabling dynamic scaling during attacks.

    \item \textbf{Stateful Security Monitoring}: Laraba et al. \cite{laraba2020defeating} modeled security monitoring functions as Extended Finite State Machines (EFSM) in P4. This approach enables detection of protocol abuse, such as ECN protocol misuse and optimistic ACK attacks, by mapping protocol states and transitions to P4 primitives.

    \item \textbf{5G Network Slice Protection}: Bonfim et al. \cite{bonfim2020realtime} developed FrameRTP4, a system for detecting and mitigating attacks in 5G network slices in real-time, using data structures like Bloom Filters, Count-Min Sketch, and Invertible Bloom Lookup Tables for efficient monitoring.
\end{itemize}

These security mechanisms demonstrate the versatility of programmable data planes; quite complicated logic can be implemented using the limited resources on a switch, in conjunction with proper algorithms and data structures. However the question remains whether pure dataplane programming is powerful enough to perform next-generation firewalling, i.e. Deep Packet Inspection. Surprisingly, it can. 

The key breakthrough enabling string and regex matching in the data plane, despite the parser's limitations, is a technique known as recirculate-and-truncate. Since the P4 parser is single-pass and cannot loop to parse arbitrary data (as seen in packet payloads!) this approach instead recirculates a packet through the switch pipeline multiple times. On each pass, the first byte of the payload is removed for inspection, allowing the switch to effectively simulate a sliding-window traversal of the packet one byte at a time.

By coupling this mechanism with a deterministic finite automaton (DFA) implemented in the match-action tables, the switch can track pattern matches across passes. Each stage in the pipeline corresponds to a transition in the DFA, enabling deep packet inspection at line rate, despite the absence of loops or complex parsing logic. This technique powers systems such as PPS, DeeP4R, and BOLT.
\begin{itemize}[leftmargin=*]
    \item \textbf{PPS} Jepsen et al.'s PPS (P4 Programmable String search) system~\cite{jepsen2019fast} was the first paper to match strings using a deterministic finite automaton (DFA) mapped to the match-action pipeline of a PISA switch. PPS stored the DFA transitions into multiple match stages in the pipeline, each performing a single-character match and state transition. The DFA is stored in a series of ternary match tables, and each table maps an input character and current state to a next state. This mapping is highly efficient (due to the fixed pipeline architecture of PISA), though the memory overhead grows with the number of DFA states and input symbols. PPS supports thousands of keyword patterns and achieves line-rate processing at 10–100 Gbps, significantly outperforming conventional CPU- and FPGA-based DPI implementations in latency and throughput.

    \item \textbf{DeeP4R} Gupta et al. extend this idea with DeeP4R~\cite{gosain2023deepp4r}, a full next-gen firewall in P4. DeeP4R implements Deep Packet Inspection by recirculating a cloned copy of the packet through the switch pipeline. On each iteration, the first byte of the clone packet is cut off for a DFA-based pattern match. The original remains untouched, and is forwarded if the clone matches no known patterns in the firewall (otherwise, it is discarded). Thus P4's \texttt{clone} and \texttt{recirculate} primitives are used to achieve byte-wise inspection, with as many passes as needed.  DeeP4R handles thousands of patterns with limited register and pipeline resource usage, and can enforce layer-7 security policies (e.g., URL blacklists) with low overhead in real-time traffic.

    \item \textbf{BOLT} Large rule sets or complex pattern collections can be too costly to match (DFA state explosion).  BOLT~\cite{zhang2021bolt} achieves more scalable multi-pattern string matching, using two optimizations. BOLT reduces DFA transitions using $k$-stride (multiple input bytes are consumed at a step) to reduce pipeline depth, and also reduces the size of the DFA using bit slicing to compactly represent the transition logic.  This achieves a balance between memory usage and matching speed, and allows for the best operation within the strict memory and stage constraints of real switches.
\end{itemize}

%%%%%%%%%%%%%%%%%%%%%%

\subsection{Cryptographic Primitives in the Data Plane}

The first challenge when considering secure connections in the data plane, is what secure operations can be supported on a programmable switch. Despite the limited instruction set of P4, several research efforts have successfully implemented cryptographic functions:
\begin{itemize}[leftmargin=*]
    \item \textbf{AES Encryption}: Chen \cite{chen2020implementing} proposed  ``scrambled lookup tables" to implement AES encryption entirely in the data plane without controller interaction. The approach supports all three variants of AES (AES-128, AES-192, and AES-256) and leverages packet recirculation to simulate the multiple rounds required by the algorithm. Evaluations showed throughput of 10.92, 8.76, and 7.37 Gbps for AES-128, AES-192, and AES-256, respectively.

    \item \textbf{Diffie-Helman with AES}: Oliveira et al \cite{dh-aes} build on the above system to include Diffie-Hellman key exchange, so the smart switch is not only performing encryption but also key setup.

    \item \textbf{Content Permutation}: Lin et al. \cite{lin2019enhancing} developed a secret permutation mechanism for P4 switches to protect 5G packet payloads. The algorithm partitions the payload into codewords and shuffles them across multiple stages of the ingress pipeline, achieving permutation at line rate.
\end{itemize}

In addition to the above ``pure'' implementations, some noteworthy implementations have made use of non-standard hardware.
\begin{itemize}[leftmargin=*]
    \item \textbf{Cryptographic Hash Functions}: Scholz et al. \cite{scholz2019cryptographic} extended multiple P4 targets to support cryptographic hash functions, including SipHash-2-4, Poly1305-AES, BLAKE2b, HMAC-SHA256, and HMAC-SHA512. Different implementations were tested on various targets including CPU-based (t4p4s), NPU-based, and FPGA-based platforms. CPU targets were easily extensible but had higher latency; NPU offered the highest throughput; and FPGA targets provided the lowest latency.
    \item \textbf{FPGA-accelerated Cryptography}: Malina et al. \cite{malina2020hardware} accelerated cryptographic operations for FPGA-based network cards and integrated them as P4 externs. They implemented symmetric ciphers (AES-GCM-256), digital signatures (EdDSA), and hash functions (SHA3) using VHDL, achieving 26.24 Gbps for cryptographic functions and 4.51 Gbps for hash functions.
\end{itemize}

\subsection{Secure Protocols in the Data Plane}

Besides cryptographic primitives, researchers have implemented complete security protocols on programmable switches. We note that so far none of these were implemented purely in the data plane.
\begin{itemize}[leftmargin=*]
    \item \textbf{P4-MACsec}: Hauser et al. \cite{hauser2020p4macsec} implemented Media Access Control security (MACsec) using P4. MACsec provides point-to-point security between enabled endpoints within a local area network. Their implementation features a two-tier control plane, with local controllers handling time-sensitive operations and a central controller for coordination. For encryption and authentication, P4-MACsec uses AES in Galois/Counter Mode (AES-GCM) implemented as P4 externs.

    \item \textbf{P4-IPsec}: The same group \cite{hauser2019p4ipsec} further implement Internet Protocol Security (IPsec), specifically the Encapsulating Security Payload (ESP) protocol in tunnel mode, with AES Counter Mode (AES-CTR) and NULL ciphers. Instead of using the Internet Key Exchange (IKE) protocol, P4-IPsec relies on an SDN controller for Security Association (SA) management, reducing message exchanges. The crypto manager is implemented either on the switch's CPU via PCIe or on an external crypto host.

    \item \textbf{P4NIS}: Li et al. \cite{liu2021softwarized} proposed a Network Immune Scheme against eavesdropping attacks in IoT networks. P4NIS implements multiple defense layers: traffic distribution across different network paths to confuse eavesdroppers, transport layer header encryption, and packet payload encryption. While the data plane handles traffic distribution and some encryption, complex encryption operations are delegated to the controller or IoT devices.
\end{itemize}

In other words, the restricted instruction set of P4 has made it impractical to implement cryptographic systems purely in the data plane with standard switches. Researchers have addressed this by using external resources (CPU, controller), and this is a limitation creating potential bottlenecks and security vulnerabilities in the communication channel. Also the actual encryption is handled using dedicated hardware and not a standard switch. We therefore identify an important research gap: to build on the early work of Chen~\cite{chen2020implementing} and create a complete security solution (such as an IPSec endpoint).

We note that there is a tension between the amount of on-chip memory needed for the look-up tables and the security of the solution (more complex cryptographic algorithms will need more memory); similarly, stronger security typically requires more complex algorithms and will reduce throughput. Another concern is whether we allow controller dependence for operations such as key management. Despite these challenges and caveats, a successful implementation of cryptography directly on standard switches would offer significant advantages for network security, enabling high-speed encryption and authentication without dedicated security appliances.

\section{Smart Switches, In-Network Computation and Machine Learning}
\label{sec:fil2}

In the previous sections, we considered the use of programmable switches in their role as switches and as middleboxes for network security. However, the increasing sophistication of cyber threats and the growing complexity of modern networks have made traditional rule-based security approaches insufficient. This section focuses on Machine Learning-based security applications within programmable networks, and on how it is possible to do at least some limited ML inference in the data plane.

ML models deployed in programmable user planes have been effective in several  security domains, including DDoS and intrusion detection. More comprehensively,
\begin{itemize}[leftmargin=*]
    \item \textbf{Intrusion Detection Systems (IDS):} ML-based IDS implemented in the user plane, like Planter~\cite{zheng2021planter} and Soter~\cite{xie2022soter}, identify malicious traffic by analyzing packet-level features directly in the switch pipeline. Compared to traditional control-plane IDS, these user-plane approaches reduce detection latency from milliseconds to nanoseconds, significantly limiting the impact of attacks.

    \item \textbf{DDoS Attack Mitigation:} Tree-based models embedded in programmable switches have proven effective for rapid Distributed Denial of Service (DDoS) detection. BACKORDERS~\cite{coelho2022backorders} implements Random Forest (RF) models in the programmable data plane to identify DDoS attacks with high accuracy. Similarly, pHeavy~\cite{zhang2021pheavy} enables detection of heavy flows characteristic of volumetric attacks through stateful flow-level monitoring in the switch.

    \item \textbf{Encrypted Traffic Analysis:} Recent advancements enable security analysis of encrypted traffic without breaking encryption. Flowrest-based approaches~\cite{akem2024encrypted} for encrypted traffic classification use statistics of packet sizes and inter-arrival times  to identify potentially malicious encrypted communications at line rate, achieving accuracy levels around 87-95\% across various encrypted application classification tasks.

    \item \textbf{Smart Grid Protection:} Ultra-low latency cyberattack detection for Smart Grid (SG) systems has been demonstrated using Decision Tree (DT) models embedded in programmable switches~\cite{akem2024ultralow}. This approach achieves 99\% accuracy in detecting attacks against Distributed Network Protocol 3 (DNP3) for industrial control systems, with sub-microsecond latency -- 17,000 times faster than control-plane solutions.

    \item \textbf{Botnet and Malware Detection:} In-Network Classification (INC)~\cite{friday2022inc} enables line-rate detection of botnet propagation in programmable switches. Similarly, Friday et al.~\cite{friday2022learning} proposed a methodology for ransomware detection in P4 switches. These approaches typically identify command-and-control communication patterns and anomalous data exfiltration characteristic of malware. 
\end{itemize}

\subsection{Architectural Approaches to ML-based Security}

Machine learning models in programmable networks can operate at three levels: per-packet, per-flow, or with joint inference that combines both. Each inference strategy has distinct tradeoffs in terms of latency, resource usage, and detection power.

\subsubsection{Packet-Level Inference}

Packet-level (PL) inference extracts features from individual packets in isolation, typically focusing on header fields. This design enables rapid threat detection without requiring flow state.

\begin{itemize}[leftmargin=*]
    \item \textbf{Comprehensive Coverage:} Every packet is inspected and classified at line rate, ensuring that malicious traffic is not missed—crucial for detecting zero-day exploits or malformed headers.
    \item \textbf{Low Latency:} Detection occurs immediately from the first packet, avoiding delays associated with flow aggregation.
    \item \textbf{Minimal State Overhead:} No per-flow state is required, conserving limited memory resources.
\end{itemize}

Several systems have demonstrated the feasibility and effectiveness of PL inference in P4 switches, including IIsy~\cite{xiong2019switches}, Planter~\cite{zheng2021planter}, Mousika~\cite{xie2022mousika}, and Soter~\cite{xie2022soter}.

However, this approach cannot capture behavioral patterns that span multiple packets. Additionally, matching against a large number of patterns for every packet can create computational bottlenecks in constrained environments.

To address these challenges, Henna~\cite{akem2022henna} introduced hierarchical packet-level inference. By decomposing complex classification tasks into simpler, cascaded subtasks, Henna achieved a 21\% F1-score improvement on a 21-category IoT identification task while maintaining efficient resource usage.

\subsubsection{Flow-Level Inference}

Flow-level (FL) inference aggregates features across multiple packets within a flow, allowing richer behavioral analysis and more accurate threat detection.

\begin{itemize}[leftmargin=*]
    \item \textbf{Behavioral Visibility:} Enables detection of attacks that unfold over time, such as data exfiltration, command-and-control, or covert channel activity.
    \item \textbf{Reduced Redundancy:} Classifying flows, rather than every packet, reduces repetitive processing.
\end{itemize}

Notable frameworks that support FL inference in programmable switches include Flowrest~\cite{akem2023flowrest}, FlowLens~\cite{barradas2021flowlens}, pForest~\cite{busse2019pforest}, and SwitchTree~\cite{lee2020switchtree}.

Despite their advantages, FL approaches introduce new challenges:
\begin{itemize}[leftmargin=*]
    \item \textbf{Delayed Classification:} Threats can go undetected during the early phase of a flow, before enough data is available.
    \item \textbf{State Exhaustion Risk:} Maintaining per-flow statistics can deplete limited memory resources and may be abused by attackers to launch state exhaustion attacks.
\end{itemize}

\subsubsection{Combined Inference}

Joint inference approaches seek to combine the responsiveness of PL models with the context-awareness of FL models.

\textbf{Sequential Multi-Model Approaches:}  
NetBeacon~\cite{zhou2023efficient} uses a series of decision tree classifiers applied at different points in the packet lifecycle. Early packets are classified using packet-level features, while later packets are matched using accumulated flow-level context. This architecture improves coverage and accuracy but increases memory and pipeline complexity.

\textbf{Unified Model Approaches:}  
Jewel~\cite{akem2024jewel} adopts a single random forest model trained on both PL and FL features. It uses dynamic feature weighting during training to adjust inference depending on packet position within a flow. Jewel outperforms both packet-only and flow-only models, achieving an average accuracy improvement of 3.2\%, while maintaining lower resource usage than NetBeacon’s multi-model pipeline.

\subsection{Advantage: Performance and Accuracy}

A standard issue in conventional security architectures is the delay in processing and communication:
\begin{enumerate}
    \item \textbf{Control-Plane ML:} Traditional SDN security approaches where ML models run in the controller induce delays of 1-100+ milliseconds due to control-user plane communication overhead~\cite{he2015measuring}.
    
    \item \textbf{External Security Appliances:} Traffic redirection to dedicated security appliances can add 1-10+ milliseconds of latency, creating a substantial window of vulnerability.
    
    \item \textbf{Hybrid Approaches:} Even solutions that distribute tasks between planes, using the user plane for feature extraction and the control plane for inference~\cite{barradas2021flowlens,amado2023poster,amado2024peregrine,seufert2024marina}, still incur millisecond-level delays.
\end{enumerate}

In contrast, user-plane ML can enable detection and mitigation at line rate, and experimental studies of in-switch security solutions demonstrate dramatic speedups compared to conventional approaches.
\begin{itemize}[leftmargin=*]
    \item \textbf{Smart Grid Security:} A DT model for attack detection in Smart Grid systems achieved a detection latency of just 356 nanoseconds -- 6,000 times faster than solutions using user-plane feature extraction with control-plane inference, and 17,000 times faster than full control-plane approaches~\cite{akem2024ultralow}.
    
    \item \textbf{DDoS Detection:} In-switch models for DDoS detection can identify and mitigate attacks within 400-500 nanoseconds, compared to several milliseconds for control-plane solutions~\cite{coelho2022backorders,zhang2021pheavy}.
    
    \item \textbf{Flow-Based Attacks:} Flowrest demonstrated detection of malicious flows with packet-processing latency below 450 nanoseconds across multiple security-relevant datasets~\cite{akem2023flowrest}.
\end{itemize}

This ultra-low latency detection can enable early responses from the IDS or firewall (as mentioned in the previous section), potentially preventing attacks from ever reaching their targets rather than merely limiting their impact after the attack.

\subsection{Challenges: Resource Constraints}

Implementing ML-based security in the data plane presents familiar technical challenges, owing to the limitations of the platform: the limited range of operations available, and the limited memory to hold look-up tables.
\begin{itemize}[leftmargin=*]
    \item \textbf{Feature Availability:} While comprehensive feature sets improve security detection accuracy, many desirable security features are challenging to implement in switches. As discussed in the previous section, payload inspection is not natively supported in P4, so if we want the ability to detect application-layer attacks we have to use complex and slow workarounds (recirculate-and-truncate). Similarly, complex temporal patterns used in security analytics are difficult to capture within switch constraints.

    \item \textbf{Resource Constraints vs. Model Complexity:} Security tasks can require complex models to detect sophisticated threats. However, switch hardware is severely resource-constrained. Commercial switches typically have only a few hundred megabytes of Static Random-Access Memory (SRAM) and a few megabytes of expensive Ternary Content-Addressable Memory (TCAM), shared between normal forwarding functions and security models.
\end{itemize}

This issue is particularly problematic when we consider that unlike general classification tasks, security models face adversaries actively attempting to bypass them. Robust and fault-tolerant defenses are particularly challenging to implement in a resource-constrained platform (for example, consider the poor performance of a neural network when heavily quantized).

One particular challenge is \textbf{zero-day detection}. Intrusion detection etc. using supervised ML, is not strong in identifying previously unseen attacks. Detecting novel attacks requires models capable of detecting anomalies rather than just known patterns, but a learning approach that can deal with such unknown patterns will likely require unsupervised or semi-supervised learning with substantial data; such models are hard to fit into a constrained environment. 

\subsection{Model Selection for Security Applications}

As programmable switches are resource-constrained, we require careful model selection for security applications. Comparative evaluation across multiple security-relevant datasets~\cite{akem2023flowrest} so far indicates that traditional machine learning, in the form of Decision Tree (DT) and Random Forest (RF) models, outperforms more complex alternatives like Neural Networks (NN) and Binarized Neural Networks (BNN) for in-switch security tasks. Across intrusion detection datasets including CICIDS-2017~\cite{sharafaldin2018toward}, UNSW-NB15~\cite{moustafa2015unsw,moustafa2016evaluation}, and NSL-KDD~\cite{tavallaee2009detailed}, RF models achieved F1-scores typically in the 95-100 range, compared to 84-91 for BNNs.

This performance advantage stems from several factors:
\begin{enumerate}
    \item \textbf{Tabular Data Efficiency:} Security-relevant network features typically form tabular data, where tree-based models often outperform deep learning approaches~\cite{grinsztajn2022tree}.
    
    \item \textbf{Structural Compatibility:} The logical structure of tree-based models aligns naturally with the Match \& Action (M/A) pipeline architecture of programmable switches.
        
    \item \textbf{Resource Efficiency:} Tree-based models require primarily simple comparison operations that map efficiently to switch hardware, whereas even simplified neural networks like BNNs can quickly exhaust switch resources.
\end{enumerate}

For example, a basic BNN with just two layers (64 and 32 neurons) completely exhausts the resources of a Tofino ASIC~\cite{siracusano2018network}, but still underperforms simpler tree-based models on security classification tasks. And of course, tree-based models are far superior in another regard: they are explainable, a factor critical for security applications where the rationale for a response must be justified.

\subsection{Future Directions}

The most direct research direction involving ML-based security in programmable networks, would be to study distillation and methods to capture powerful models on a resource-constrained platform. For example, Trustee \cite{jacobs2022ai} approximates a neural network with a best-fit decision tree. The challenge is to get models small enough to fit on a switch, but good enough to compete with machine learning models on the controller working in concert with the switch \cite{doriguzzi2024p4ddle}.

We note that future network security solutions could certainly benefit from hardware designed specifically to support ML-based security functions, potentially through specialized computing units integrated into programmable switches (extern logic, as seen in DeepMatch~\cite{hypolite2020deepmatch} for example). But the cost for this would be the loss of compatibility with standard hardware, and potentially the loss of real-time line-rate decisions. In the immediate future we are more hopeful that innovation on the software side, with possibly some limited support from hardware, will make it possible to build more robust high-performance models that still fit within switch resource constraints. Further open research problems include:
\begin{itemize}[leftmargin=*]
    \item \textbf{Unsupervised Anomaly Detection:} Current in-switch implementations focus primarily on supervised learning for known attack classes. Genos~\cite{li2024genos} represents an initial step toward unsupervised approaches, but more research is needed to enable effective zero-day attack detection within the constraints of a programmable switch.

    \item \textbf{Cross-Device Coordination:} We note that programmable switches have the advantage of always being part of a network.  In other words, they may be computationally poor, but are rich in communication bandwidth. Distributing the load of security inference across multiple network devices could enable more comprehensive threat detection without requiring stronger computational power from individual devices. In this special case of in-network multi-agent computation, the open problem is developing efficient coordination mechanisms for cooperative inference across devices.

    \item \textbf{Adaptive Security Models:} Mechanisms for efficiently updating in-switch security models in response to emerging threats, without disrupting network operations, would improve responsiveness to evolving attack landscapes. The open problem here is to resist catastrophic forgetting when exposed to new challenges.
\end{itemize}

\section{Discussion - Current Challenges}

Programmable switches offer significant advantages for network security, but also significant challenges as a platform. As previously noted, switches can only perform a limited number of operations per packet (owing to the need to process them in real time), and the limited instruction set (no native division or floating-point operations) makes it extremely challenging to implement the algorithms used in ML inference or cryptography. Several approaches have been tried to work around this constraint:
\begin{itemize}[leftmargin=*]
\item \textbf{Approximation}: Ding et al. \cite{ding2020estimating} proposed P4Log and P4Exp algorithms to estimate logarithms and exponential functions using only P4-supported arithmetic operations, enabling entropy calculation entirely in the data plane.

\item \textbf{Precomputation}: Values are precomputed and stored in match-action tables or registers, converting complex calculations into simple lookups.

\item \textbf{Binarized Models}: Qin et al. \cite{qin2020line} implemented Binarized Neural Networks in P4, converting complex ML operations into simpler bitwise operations supported by the switch.
\end{itemize}
But as noted in the previous sections, these approaches are not universally successful. And supplementing with external computation -- for instance, Hauser et al. \cite{hauser2019p4ipsec} offloading IPsec crypto manager functions: to the internal CPU module connected via PCIe, or to an external crypto host -- compromises the performance gain from in-network computation.

The other main challenge is the limited on-switch memory available for tables and metadata, which can be as low as tens of megabytes for some switches~\cite{edgenetworks2021wedge}. This constrains the number of flows or packets that can be tracked, the mappings for traffic label switching or re-mapping (say, for anonymization), the state information stored, and the look-up tables used to work around the limited native operations. Attempts to reduce the memory usage include:
\begin{itemize}[leftmargin=*]
    \item \textbf{Data Structures.} Data structures like Count-Min Sketch \cite{cormode2005improved}, Bloom Filters, and multi-stage hash tables help maximize the use of limited memory. For instance, Chen et al. \cite{chen2020measuring} stored TCP packet records in a hash table using memory arrays across pipeline stages.
    \item \textbf{Compressed Representations.} FlowLens \cite{barradas2021flowlens} uses a compact representation of packet distributions (flow markers) to track network behavior with minimal memory usage. Similarly, BOLT \cite{zhang2021bolt} uses optimized representation to compress automata for regular-expression matching.
\end{itemize}
One possible direction of future work could focus on increased cooperation between switches, or switches and servers, as agents performing distributed computing. For instance, Kim et al. \cite{kim2018generic} suggest accessing DRAM on data center servers through RDMA-capable NICs, allowing switches to utilize external memory with minimal latency impact (1-2 microseconds). Smart allocation of memory resources, with a hierarchy of access similar to cache/ memory/ disk, would be a natural architecture for in-fabric network computation.

\section{Conclusion}

This survey has shown that despite tight resource constraints, programmable switches can now serve as platforms for an unexpectedly wide range of high-performance security applications. From DDoS mitigation and firewalls to cryptographic primitives and machine learning inference, the network fabric itself is becoming an active layer of defense.
With this evolution, programmable switches are no longer just fast routers or packet filters, but essential building blocks for secure, autonomous, and adaptable networks. 

The advantage of this new platform is clear: switches can respond orders of magnitude faster than traditional software-based solutions, and are also flexible enough to adapt to evolving threats without hardware replacement. Their key advantages are line-rate processing, visibility into network behavior, and agility in responding to new threats. But they have their own challenges: memory constraints, limited processing capabilities, deep packet inspection limitations, and (if we rely on extensions in the hardware) interoperability concerns. 

Our aim in this paper is not only to survey the field, but also to draw attention to innovative approaches that address these challenges, from efficient data structures and approximation techniques to recirculate-and-truncate and look-up-table precomputation. With continuing advances in data structures, inference methods, and switch architectures, the prospect of self-defending, self-healing and self-managing networks is fast becoming a reality. By systematizing the state of the art, the current limitations, and current and possible future techniques for overcoming them, we aim to contribute to the realization of this vision.

\section*{Author Biographies} 
\begin{description}
    \item[Gursimran Singh] is a PhD candidate at Rochester Institute of Technology. His research focus is on network security applications in the data plane.
    \item[H.B. Acharya] is an assistant professor of computer science at Oklahoma State University. His interests are in the area of Networks and Cybersecurity, including topics like data plane network security, internet interference,etc.
    \item[Minseok Kwon] is a professor in the Department of Computer Science at Rochester Institute of Technology. His main research interests are in the area of computer networks, distributed computing, and network security.
\end{description}
% AUTHORS  

% Short Biography

\end{document}